\documentclass{aa}
\usepackage{epsfig}
\usepackage{natbib}
\usepackage{graphicx}

\usepackage{color}

\usepackage{graphics}
\usepackage{amssymb,amsmath}
\usepackage{amsmath}
\usepackage{psfrag}
\usepackage{graphicx}
\newcommand{\ve}[1]{\mathbf{q1}}

\newcommand{\f}{\frac}

\newcommand{\be}{\begin{equation}}      
\newcommand{\ee}{\end{equation}}      
      
\newcommand{\bef}{\begin{figure}}      
\newcommand{\eef}{\end{figure}}      
\newcommand{\bea}{\begin{eqnarray}}    
\newcommand{\eea}{\end{eqnarray}}      

\newcommand{\av}[1]{\ensuremath{\left\langle q1 \right\rangle}}

\def\sposeq1{\hbox to 0pt{q1\hss}}
\def\ltapprox{\mathrel{\spose{\lower 3pt\hbox{$\mathchar"218$}}
\raise 2.0pt\hbox{$\mathchar"13C$}}}
\def\gtapprox{\mathrel{\spose{\lower 3pt\hbox{$\mathchar"218$}}
\raise 2.0pt\hbox{$\mathchar"13E$}}}
\def\inapprox{\mathrel{\spose{\lower 3pt\hbox{$\mathchar"218$}}
\raise 2.0pt\hbox{$\mathchar"232$}}}

\newcommand{\tve}[1]{\tilde{\boldsymbol{q1}}}

\def\bse{\begin{subequations}}
\def\ese{\end{subequations}}

\def\lsim{\raise 0.4ex\hbox{$<$}\kern -0.8em\lower 0.62ex\hbox{$\sim$}} 
\def\gsim{\raise 0.4ex\hbox{$>$}\kern -0.7em\lower 0.62ex\hbox{$\sim$}}

\def\f0N{f_0^{(N)}}
\def\bec{\begin{center}}
\def\eec{\end{center}}

\begin{document}

\title{Angular momentum
  generation in cold gravitational collapse} 
  
\titlerunning{Angular momentum generation in cold collapse} 
  
\authorrunning{Benhaiem et al.}  
  
  \author
      {D. Benhaiem \inst{1} , M. Joyce\inst{2,3}, F. Sylos
        Labini \inst{4,1,5} and
        T. Worrakitpoonpon \inst{6}}

        \institute{Istituto dei Sistemi Complessi
        Consiglio Nazionale delle Ricerche, Via dei Taurini 19, 00185
        Rome, Italy
        \and UPMC Univ Paris 06, UMR 7585, LPNHE,
        F-75005, Paris, France
        \and CNRS IN2P3, UMR 7585, LPNHE,
        F-75005, Paris, France
         \and 
        Centro Studi e Ricerche
        Enrico Fermi, Via Panisperna 89 A, Compendio del Viminale,
        00184 Rome, Italy 
        \and INFN Unit Rome 1, Dipartimento di
        Fisica, Universit\'a di Roma Sapienza, Piazzale Aldo Moro 2,
        00185 Roma, Italy 
        \and Faculty of Science and Technology,
        Rajamangala University of Technology Suvarnabhumi, Nonthaburi
        Campus,  Nonthaburi 11000, Thailand}

\date{Received / Accepted}

\abstract{
During the violent relaxation of a self-gravitating system a
significant fraction of its mass may be ejected. If the time varying
gravitational field also breaks spherical symmetry this mass can
potentially carry angular momentum. Thus starting initial
configurations with zero angular momentum can in principle lead to a
bound virialized system with non-zero angular momentum.  We explore
here, using numerical simulations, how much angular momentum can be
generated in a virialized structure in this way, starting from
configurations of cold particles which are very close to spherically
symmetric. For initial configurations in which spherical symmetry is
broken only by the Poissonian fluctuations associated with the finite
particle number $N$, with $N$ in range $10^3$ to $10^5$, we find that
the relaxed structures have standard ``spin" parameters $\lambda \sim
10^{-3}$, and decreasing slowly with $N$.  For slightly ellipsoidal
initial conditions, in which the finite-$N$ fluctuations break the
residual reflection symmetries, we observe values $\lambda \sim
10^{-2}$, of the same order of magnitude as those reported for
elliptical galaxies.
The net angular momentum vector is typically aligned close to normal
to the major semi-axis of the triaxial relaxed structure, and also
with that of the ejected mass. This simple mechanism may provide an
alternative, or complement, to ``tidal torque theory" for
understanding the origin of angular momentum in astrophysical
structures.
}

\maketitle


\section{Introduction} 

 {Observations have shown already several decades ago that galaxies
  of all types generically have significant angular momentum.  Its
  origin remains a fascinating open theoretical problem (for a review,
  see e.g. \cite{romanowsky+fall_2012}).  Globular clusters have also
  been more recently observed to have net rotation
  \citep{globclu1,globclu3,globclu2}.  The most popular theory to
  account for angular momentum in virialized structures} is so-called
``tidal torque theory'' in which the virialized structure gains
angular momentum by the action of the torque due to the tidal fields
generated by surrounding structures \citep{Peebles_1969}.  We explore
here a distinct mechanism which, despite its simplicity, appears not
to have been considered in the literature, apart from in one recent
study of cold spherical collapse by one of us
\citep{worrakitpoonpon_2014}: generation of angular momentum by
ejection of matter in violent relaxation.  Indeed violent relaxation
of self-gravitating structures is characterized by very large
amplitude variations of the mean field potential of a structure, which
can give enough energy to particles to escape even if all the mass is
initially bound. The amount of mass ejected depends strongly on the
initial conditions, varying from zero to as much as $40 \%$ for highly
uniform and completely cold initial conditions
\citep{joyce_etal_2009}\footnote{Mergers of virialized structures can
  also lead to ejection of a significant mass, in particular when the
  two structures have very different mass \citep{carucci_etal_2014}.}.
If, in addition, the mass distribution is not spherically symmetric
during the violent phase of the relaxation, this ejected mass would be
expected to carry at least some angular momentum --- and ``generate"
in the remaining bound mass an (equal and opposite) amount of angular
momentum. In other words the two components of the mass --- the part
that is ejected, and the other part which remains bound --- can exert
a net torque on one another during the violent relaxation leading to a
net angular momentum for both of them.  Further, given that violent
relaxation starting from a wide range of cold initial conditions is
often characterized by a very strong breaking of spherical symmetry
even when the initial conditions are not --- leading in particular to
triaxial relaxed structures (e.g. \cite{merritt+aguilar_1985,
  barnes_etal_2009,Benhaiem+SylosLabini_2015,SylosLabini+Benhaiem+Joyce_2015})
--- one might expect the effect to be far from negligible.

We explore here, using numerical simulations, how much angular
momentum can be generated {in a virialized structure by this
  mechanism. More specifically we consider an isolated initially cold
  distribution of matter in open boundary conditions, and without
  expansion, and a range of initial spatial distributions} which are
very close to spherically symmetric: (i) particles distributed around
a center following a mean density profile which decays as a power law
of the radial distance, or (ii) particles distributed with uniform
mean density inside an ellipsoidal region.  These are initial
conditions which have been extensively studied in the literature (see
references below) to investigate the processes involved in the
formation of galaxies and other astrophysical structures. We note
that, in a cosmological context, the simulations can be taken to
represent the evolution, in physical coordinates, of a single isolated
overdensity with the chosen initial profile.  {On the other hand
the relation of such simulations to the more general case of the 
evolution an overdensity in an expanding universe (which cannot 
necessarily be well approximated as isolated) is a more 
complex issue, and will be discussed in detail in a 
separate forthcoming publication.
 It involves several additional parameters
which can play a role in the dynamics (e.g. how precisely
the background density is modelled) and additional
numerical issues (notably concerning the control of
numerical accuracy without energy conservation, see
 \cite{jsl13,sl13}).}

In the present context
the fluctuations breaking spherical symmetry in the initial conditions
are evidently of central importance for the phenomenon we are
studying.  In the first case spherical symmetry is broken only by the
finite particle number fluctuations, while in the second case these
fluctuations break the residual reflection symmetries.We find that
almost all these initial conditions --- except where there is
negligible mass ejection --- indeed lead to a measurable net angular
momentum in the relaxed virialized structure. The magnitude of this
angular momentum is larger by about an order of magnitude in the
latter class, and in many cases is sufficiently large to suggest that
the mechanism could potentially account for angular momenta observed
in astrophysical structures. Indeed, the typical measured values of
the spin parameters of galaxies (see
e.g. \cite{hernandez2007empirical})
are of the same order of magnitude as those we find in our simulations. 

The paper is organized as follows. We first present the details of the
numerical simulations and initial conditions in Sect.\ref{numsim}.
{In the following section we then present our results, first
  describing the relevant features of the evolution qualitatively and
  then giving the quantitative results. In Sect.\ref{diss_concl} we
  summarize our results and conclude.}

\section{Numerical simulations} 
\label{numsim}

\subsection{Initial conditions} 

In detail the initial conditions of which we study the evolution are
the following:

\begin{itemize}

\item $N$ particles distributed {\it randomly} inside a sphere of
  radius $R_0$, following the radius-dependent density profile $\rho
  (r) \propto r^{-\alpha}$ where $r$ is the radial distance from the
  cent re and $\alpha$ is a constant.  We will refer to these as
  ``spherical initial conditions''.  The particles have vanishing
  initial velocities. We report here results for the range of $\alpha$
  with $0\leq \alpha \leq 2$.  We restrict to this range as in a
  previous study
  \citep{syloslabini_2012,SylosLabini+Benhaiem+Joyce_2015} we have
  observed that, for $\alpha >2$, there is negligible ($\ll 1 \%$)
  mass ejection. As shown in this same study, the evolution leads,
  except for $\alpha$ very close to zero, to virial equilibria which
  are very non-spherically symmetric, and typically triaxial.  We have
  varied the number of particles $N$ from $10^3$ to $10^5$.

\item $N$ particles are distributed randomly in a prolate ellipsoidal
  region. We define $a_3$ to be the largest semi-principal axis and
  $a_1=a_2$ the smaller ones.  The three eigenvalues of the inertia
  tensor are simply
\be
\label{lambda} 
\Lambda_i = \frac{1}{5} M (a^2_j+a^2_k)
\ee
where $M$ is the total mass of the system, and $i \ne j \ne k$ and
$i,j,k=1,..,3$: from the definition of the semi-principal axes we have
$\Lambda_1 \ge \Lambda_2 \ge \Lambda_3$. Note that the principal axis
corresponding to the eigenvalue ${\Lambda}_1$ is oriented in the
direction of the shortest semi-principal axis $a_1$, while the
principal axis associated with ${\Lambda}_3$ is in the direction of
the longest one $a_3$. The particle number $N$ spans again the range
from $10^3$ to $10^5$.

The shape at any time may then be characterized by the
parameter~\footnote{Note that these parameters are generally defined
 as a function of the semi-principal axes $a_1,a_2,a_3$ rather than
 as a function of the eigenvalues (see Eq.\ref{lambda}). However for
 small deformations of a perfect sphere, which is the case we
 consider here, these definitions are almost equivalent.  The
 advantage of this definition of the parameters using the eigenvalues
is that it can be used to characterize any distribution.}
\be 
\label{iota} 
\iota(t)= \frac{\Lambda_1(t)}{\Lambda_3(t)} -1 \;. 
\ee
We will refer to these as ``ellipsoidal initial
conditions", and report here results for value of $\iota$ in the range
from $\iota(0)=0.01$ to $\iota(0)=0.25$. The mass ejection and amplification 
of the spherical symmetry breaking during the evolution from these
initial conditions has been studied in \cite{Benhaiem+SylosLabini_2015}.
\end{itemize}

\subsection{Numerical simulations} 

We have used the N-body code {\tt Gadget} \citep{springel_2005}.  All
results presented here are for simulations in which energy is
conserved to within {\it one tenth of a percent} over the time scale
evolved, with maximal deviations at any time of less than half a
percent (see
\cite{Benhaiem+SylosLabini_2015,SylosLabini+Benhaiem+Joyce_2015} for
more details).  This accuracy has been attained using values of the
essential numerical parameters in the {\tt Gadget} code [$0.025$ for
  the $\eta$ parameter controlling the time-step, and a force accuracy
  of $\alpha_F= 0.001$] in the range of typically used values for this
code. In the specific cases of spherical initial conditions with
$\alpha=0$, which is singular in the limit $N \rightarrow \infty$, the
treatment is as detailed in \cite{joyce_etal_2009} (see also
\cite{worrakitpoonpon_2014}).  We have also performed extensive 
tests of the effect of varying the force smoothing parameter 
$\varepsilon$,  and found that we obtain very stable results 
provided it is significantly smaller than the minimal size reached 
by the whole structure during collapse \footnote{More specifically 
we have found very stable results for $\varepsilon$ in the range 
$R_g^{min}/ \varepsilon \in [10,200]$ where  $R_g^{min}$ is the 
minimal value of the gravitational radius (defined further below)
reached  during the collapse.}. We will discuss the dependence on 
particle number $N$ of our results below.

As noted above, our simulations are performed with open boundary
conditions and in a non-expanding background, but can be taken to
represent well the evolution of an isolated overdensity in an expanding 
universe, {which, at the initial time, is of high density compared 
to the mean mass density of the universe and at rest, in physical 
or comoving coordinates}.
In this respect we could perform the simulations here using the
expanding universe version of the code and periodic boundary 
conditions. {The size of the periodic box relative to
that of the initial sphere would then fix the relative overdensity 
represented by the sphere at the initial time (and thus also 
the time scale for virialization compared to the Hubble time)}. 
Such a simulation {gives, as discussed in detail
and studied at length in \cite{jsl13,sl13},  identical results in physical 
coordinates, modulo finite size corrections suppressed by the 
ratio of the size of the structure to the size of the periodic box,  
and possible numerical effects (including the effect of smoothing)}.  
Indeed, as discussed in  \cite{jsl13,sl13}, the differences 
between such simulations and those in open non-expanding
space provides information about such numerical effects, and 
indeed a tool to control better expanding universe simulations.  
Our ``direct" simulations in physical coordinates are preferable 
because they are simpler and, notably, far easier to control 
for accuracy (via energy conservation). 

Given that we are interested here in the angular momentum, we
also test our simulations for its overall conservation and, as
detailed below, compare the measured angular momentum of the final
bound structure with the numerical error. We will also check
systematically for the accuracy of conservation of the total
linear momentum.

\section{Results} 
\label{results}

\subsection{Qualitative description of mechanism} 

Numerical studies of evolution from some of these initial conditions,
or very similar ones with small but non-zero virial ratios, have been
reported extensively in the literature  (see e.g. \cite{Henon_1973,vanalbada_1982,Aarseth_etal_1988,
theis+spurzem_1999,Boily_etal_2002, joyce_etal_2009,syloslabini_2012,syloslabini_2013,Benhaiem+SylosLabini_2015,SylosLabini+Benhaiem+Joyce_2015}),
with an emphasis in particular on the study of the shape and profile
of the virialized structure. Because the system is initially cold it
undergoes in all cases a strong collapse, in a time of order
$1/\sqrt{G\rho_0}$ where $\rho_0$ is the initial mean density,
followed by a re-expansion which leads rapidly to virialization of
most of the mass.  The ``degree of violence'' of the collapse ---
which can be characterized roughly by the maximal contraction the
system undergoes --- varies with the initial condition. The most
extreme case is the spherical case ($\alpha=0$) in which the collapse
is singular in the limit $N \rightarrow \infty$.

While both early studies, and most studies since then, have focused on 
the properties of the virialized structure (e.g. its density profile, shape, and 
velocity distributions) the phenomenon of mass (and energy) ejection has been  
considered in detail only more recently: for the case of
cold uniform spherical conditions (i.e. the case $\alpha=0$
here) in \cite{joyce_etal_2009}, for the case $\alpha=0$ with non-zero
initial velocities in  \cite{syloslabini_2012} 
and for ellipsoidal initial conditions  in \cite{Benhaiem+SylosLabini_2015}
\footnote{{Mass ejection in mergers of two virialized structures is
    studied in detail in \cite{carucci_etal_2014, samsing_2015}.}}.
Essentially the amount of mass ejected depends on the degree of
violence of the relaxation, which can be quantified roughly by the
maximal contraction attained by the system during the collapse
phase. The reason for this correlation between the ejection of mass
and the violence of collapse becomes easy to understand when the
mechanism for ejection is studied
\citep{joyce_etal_2009,carucci_etal_2014,Benhaiem+SylosLabini_2015,SylosLabini+Benhaiem+Joyce_2015}:
The ejected particles are essentially those of which the fall times to
the center of the structure during collapse are longest. As a
consequence they pass through the center of the whole structure when
it has begun re-expanding, which means that they travel into a deeper
potential than the one that they travel out of. As a result they pick
up an energy ``kick''. The magnitude of this energy gain is directly
related to the strength of the potential at this time, which depends
essentially on the extent of the contraction.

For the generation of angular momentum, the second crucial ingredient, 
in addition to mass ejection, is the breaking of spherical symmetry.
Indeed mass ejection can occur in a purely radial gravitational field,
but will not then generate angular momentum.  That cold initial 
conditions in the class we are considering lead to virialized states 
that are very far from spherically symmetric was observed first 
by \cite{merritt+aguilar_1985}, and extensively studied in the 
literature
(see e.g. \cite{aguilar+merritt_1990,theis+spurzem_1999,boily+athanassoula_2006,barnes_etal_2009}).
This instability of initially spherical systems to relaxation to
non-spherical viral equilibria is usually referred to as ``radial
orbit instability'', because of its apparent link to an
instability of equilibrium systems with purely radial orbits
originally proved by \cite{antonov_1961,Fridman+Polyachenko_etal_1984}.  
In two recent papers we have studied in detail the development of this
  asymmetry during the collapse phase for both the classes of initial
  conditions we study in this paper (spherical initial conditions in
  \cite{SylosLabini+Benhaiem+Joyce_2015}, and ellipsoidal initial
  conditions in \cite{Benhaiem+SylosLabini_2015}).
These studies reveal that the same process involved in the energy
ejection leading to mass ejection plays a crucial role, for many
initial conditions, in amplifying the symmetry breaking.  Indeed,
amongst the late arriving particles, there is also a spread in arrival
times as a function of angle, which leads to an energy injection, and
thus a spatial distribution of mass, which is also a function of
angle. The effect is maximal in enhancing the final asymmetry in the
range $\alpha \in [0.5, 1.5]$. It is, on the other hand, relatively
suppressed in the limit $\alpha=0$ because the particles' fall times,
and consequently their energy changes, are much less strongly
correlated with their initial radial position in this case, leading to
a ``washing out" of the effect. The development of the asymmetry in
the mass distribution, and thus in the gravitational potential, is the
combined result of the growth of the initial finite $N$ fluctuations
breaking spherical symmetry through gravitational instability ---
known in this case as the \cite{Lin_Mestel_Shu_1965} 
instability --- which is then amplified by the energy ejection
to particles which also leads to mass ejection.

\begin{figure}
\vspace{1cm}
{
\par\centering \resizebox*{9cm}{7cm}{\includegraphics*{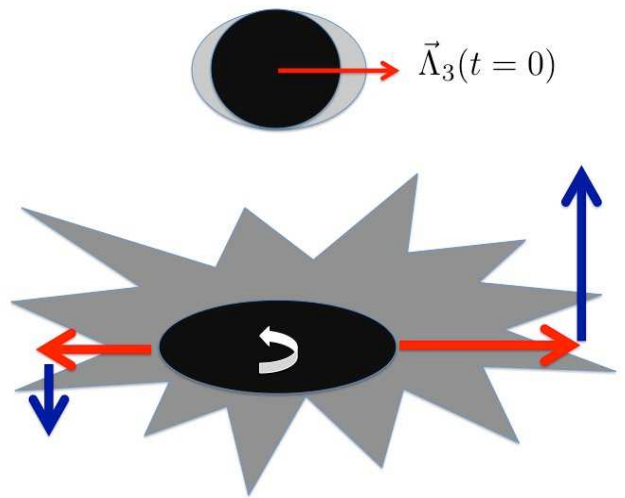}
}
\par\centering
}
\caption{Schematic representation of the generation of angular
  momentum in the ejected particles.  The upper panel shows
  schematically symmetry breaking in the initial cloud. When particles
  are ejected both their radial and traverse components will generally
  have some dependence on angle, leading to a net angular momentum
  being carried away by the ejected particles. }
\label{inertia_toy}
\end{figure}

A schematic representation of the mechanism is given in Figure 
\ref{inertia_toy}.  Any given realization of our 
initial conditions has a longest semi-principal axis, along
which particles are on average further from the orthogonal
plane. In the case of the spherical initial conditions the
non-zero effective ellipticity is a finite $N$ effect, while in the 
ellipsoidal initial conditions it is dialed by the parameter $\iota(0)$.
In the first phase of collapse this initial asymmetry is 
amplified by the (gravitational) instability of
\cite{Lin_Mestel_Shu_1965}  with the collapse occurring latest along 
the longest axis.  Particles
 arriving from the corresponding directions pass through the center as
 the structure is already re-expanding in the other directions,
 leading to a greater energy injection to them. As the potential they
 are traveling is not spherically symmetric the particles also gain
 traverse velocities and thus angular momenta with respect to the
 center. As there are also fluctuations breaking 
 rotational symmetry, which grow in the course of the collapse, both
 the radial and transverse velocities will vary as a function of
 direction, and, for example, those ejected in opposite directions
 will have slightly different angular momentum.  The ejected outgoing
 particles can then carry net non-zero angular momentum leaving behind
 the opposite angular momentum in the particles which virialize in the
 bound structure.

\subsection{Measurement of angular momentum} 

We now focus on the evolution of the angular momentum during the
relaxation. More specifically we decompose the total angular momentum
$\vec{L}_T$ at any time into two components:
\be
\label{L_total1} 
\vec{L}_T 
=  \vec{L}_b + \vec{L}_f\,
\ee
where $ \vec{L}_b$ ($\vec{L}_f$) is the total angular momentum of the
``bound" (``free") particles, i.e., which have negative (positive)
energy at the given time. In practice particles' energies almost never
change sign more than once, i.e., the asymptotically ejected particles
are those which are tagged as ``free" when their energy becomes
positive.  Further we consider the decomposition of $\vec{L}_b$ as
\be
\label{L_total} 
\vec{L}_b=  \vec{L}_b^{com} + \vec{L}_b^p\,
\ee
where $\vec{L}_b^{com} = M_{b} \vec{R}_b \times \vec{V}_b$ is the
angular momentum of the center of mass, located at $\vec{R}_b$ and
moving with velocity $\vec{V}_b$, of the bound particles, and
$\vec{L}_b^p$ is the angular momentum of the bound particles with
respect to their center of mass. It is the latter which interests us
here primarily, but we also monitor $\vec{L}_b^{com}$ and $\vec{L}_T$,
measured with respect to the fixed origin at the center of mass of the
initial configuration, to assess the accuracy of the simulation for
which the total angular momentum should be conserved (and equal to
zero).

We will measure time in units of 
\be
\label{tauc}
\tau_c=\sqrt{ \frac { 3 \pi} {32 G
    \overline{\rho}_0}}
\ee
where $\overline{\rho}_0$ is the total initial mass density, i.e., the
total mass divided by the volume of the initial sphere or ellipsoid.
It corresponds to the time for a particle initially at the outer
periphery of a cold sphere with this mass density to fall to
its center in the continuum approximation, (i.e. taking $N \rightarrow
\infty$ and keeping the initial mass density profile fixed) and
without shell crossing.

We will measure angular momentum in the natural units given by
\be
L_0= \frac{ GM^{5/2} } {\sqrt{|E_0|}} 
\label{normalisation of L}
\ee
where $M$ is the total (initial) mass of the system, and $E_0$ its
total (initial) energy (equal to its potential energy). Further, we
will report results for the angular momentum of the bound mass
$\vec{L}_b^p$ given in terms of the so-called ``spin parameter''
$\lambda$ as defined by \cite{Peebles_1969,knebe2008}:
\be \lambda = \frac{|\vec{L}_b^p|}{L_0^b} \ee
where ${L_0^b}$ is 
\be
L_0^b= \frac{GM_b^{5/2} } {\sqrt{|E_b|}} 
\label{normalisation of L2}
\ee 
where $M_b$ is the bound mass and $E_b$ ($W_b$) the total
  (gravitational potential) energy of this mass (with respect to its
  center of mass)\footnote{Another commonly used normalization for
  the spin is that introduced in \cite{Bullock_2001}, defined by
  $\lambda^\prime = |\vec{L}_b^p|/L_1$ with $L_1=\sqrt{2M_b^3GR_b}$
  with $R_b=-\frac{3GM_b^2}{5W_b}$ where $W_b$ is the potential
  energy.  For a virialized structure, with $E_b=W_b/2$,
  $\lambda^\prime =\sqrt{5/3} \lambda$.}.

\subsection{Results for a chosen initial condition}

\begin{figure}
\vspace{1cm}
{
\par\centering \resizebox*{9cm}{8cm}{\includegraphics*{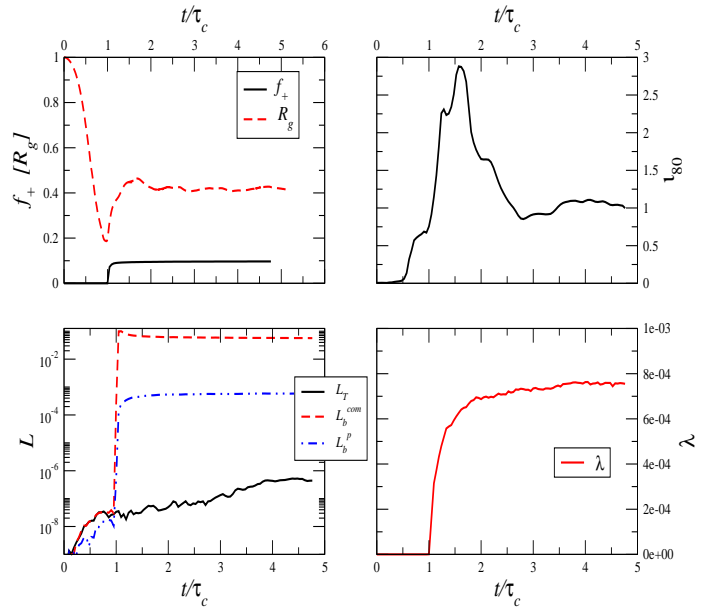}}
\par\centering
}
\caption{Evolution from spherical initial condition with $\alpha=1$,
and $N=10^5$ : (top left) the fraction of particles with positive
energy $f_+$ (full black line) and the gravitational radius $R_g$
(dotted red line) in units of the initial sphere radius $R_0$ ; (top
  right) the flatness ratio $\iota_{80}$ for the $80\%$ most bounded
  particles; (bottom left) the total angular momentum $L_T$ (full
  black line) with respect to the initial center of mass, the angular
  momentum of the center of mass of the bound particles $L_b^{com}$
  (red dashed line) and the angular momentum of the bound particles
  $L_b^{p}$ (blue dashed-dotted line) with respect to their center of
  mass (all in units of $L_0$ defined in Eq. \ref{normalisation of L}); 
 (bottom right) the spin parameter $\lambda$.}
\label{Fig1_alpha1_N1e5} 
\end{figure}

We present results first for the case of spherical initial conditions
with $\alpha=1$, and $N=10^5$ particles. We consider this case as its
evolution is typical of what we observe for all our different initial
conditions.  The panels of Fig.\ref{Fig1_alpha1_N1e5} show for this
case, as a function of time, (i) the fraction $f_+$ of the initial
mass in free particles, and the gravitational radius defined
as $R_g(t)  = \frac{G M_b^2(t)}{|W_b(t)|}$, (ii) the
flattening ratio $\iota_{80}$ {as defined by (\ref{iota}), with the
subscript indicating that the inertia tensor is calculated using the
particles with energy less than 20 \% of the energy of the most
bound particle}, (iii) the magnitude of the angular momenta
  $|\vec{L}_T|$, $|\vec{L}_b^{com}|$, $|\vec{L}_b^p|$ (on a
  logarithmic scale), and (iv) the spin parameter $\lambda$.

The first plot illustrates clearly, as described above, that at
$\tau_c\approx 1$, the system reaches its maximal contraction, and the
strong energy injection which leads to particle ejection occurs in a
very short interval around this time, corresponding approximately to
the dispersion in fall times of the mass.  Likewise we see in the
second plot, showing the behavior of $\iota_b$, that the {breaking of
  spherical symmetry grows monotonically from the beginning and is
  then strongly amplified} around the maximal collapse by the energy
injection (as described in detail in
\cite{SylosLabini+Benhaiem+Joyce_2015}).

As anticipated, as can be seen from the lower two plots, the bound
mass indeed goes toward a stationary state with non-zero angular
momentum (relative to its center of mass).  This final angular
momentum of the bound mass, albeit small in the natural units, is
clearly not numerical in origin: it is about three orders larger than
the final total angular momentum which gives a measure of the
violation of angular momentum conservation due to numerical
error. Further we have performed several tests varying the essential
numerical parameters and found our results to be stable.  On the other
hand the total angular momentum $\vec{L}_b$ of the bound mass (with
respect to the initial center of mass of the whole system) is
dominated by its translational motion: the bound mass ``recoils" with
a net linear momentum compensating that of the ejected mass, along an
axis which is off-set from the initial center of mass. We will give
some further details below on this ejected linear momentum, which is
measured numerically to a precision of order $10^{-6}$ (i.e. the
measured momentum change of these particles is $10^6$ times larger
than the change in the total momentum due to numerical error).

\subsection{Generation of angular momentum during collapse}

It is interesting to follow the dynamics leading to the generation of
the ejection of net angular momentum in the collapse phase.  During
this phase the mass which gets ejected has not yet undergone the
energy boost which leads to its ejection, but the gravitational
potential has already developed very significant asymmetry which leads
the particles to have non-radial velocities, and thus non-zero angular
momentum (while the total angular momentum remains zero).

\begin{figure}
\vspace{1cm}
{
\par\centering \resizebox*{9cm}{8cm}{\includegraphics*{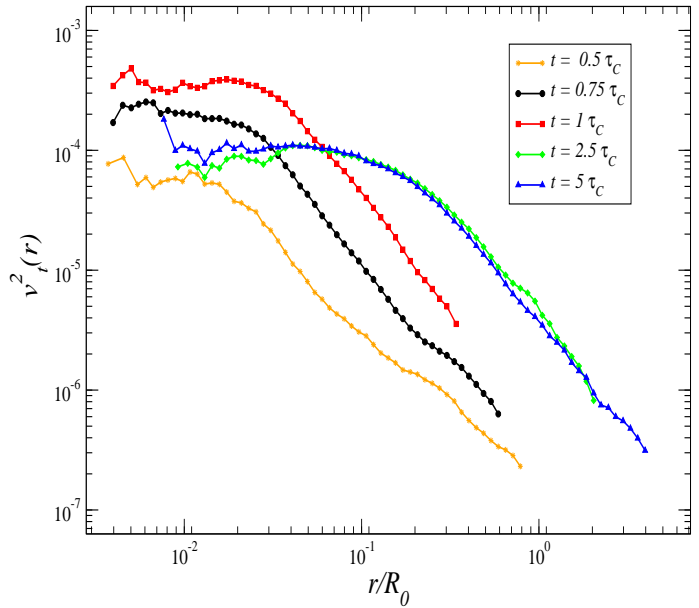}}
\par\centering
}
\caption{Average in radial shells of the square of the transverse
  component of particle velocities (in units in which  
  $v_0=\sqrt{GM/R_0}$ is unity), as a function of radius, at each
  of the indicated times, for a simulation with $N=10^5$ particles of
  spherical initial conditions with $\alpha=1$. Note that already well
  before the system reaches it maximal contraction, at $t\approx 1$,
  the Poissonian fluctuations breaking spherical symmetry have been
  amplified to produce tangential velocities larger than in the final
  virialized state.}
\label{VtVr}
\end{figure}

Shown in Fig.\ref{VtVr} is, again {for the case $\alpha=1$} and
$N=10^5$, the measured dispersion of the transverse velocities in
radial shells, as a function of radius, and at different times during
the evolution. The velocity is normalized in units of $v_0=\sqrt{GM/R_0}$ .  
We observe that the transverse velocities are already, at $t=0.5
\tau_c$ comparable with, and, at $t=0.75 \tau_c$, even larger than
their magnitude in the final virialized structure. Note that these are
the transverse velocities produced by the growth of the initial density
fluctuations breaking spherically symmetry through gravitational
instability, well before the energy injection to particles arriving
around $t\approx \tau_c$ which leads to the mass (and angular
momentum) ejection.
At longer time scales, i.e. $t > \tau_c$, the outer shell,
  corresponding to very weakly bound particles with energy very close to zero, is 
  still expanding as these particles are still traveling outward on their large
  orbits. On the other hand the inner shell has reached a stationary state, as
  can be seen by comparing the number density profiles at $t= 2.5
  \tau_c$ and at $t=5 \tau_c$.

\begin{figure}
\vspace{1cm}
{
\par\centering \resizebox*{9cm}{8cm}{\includegraphics*{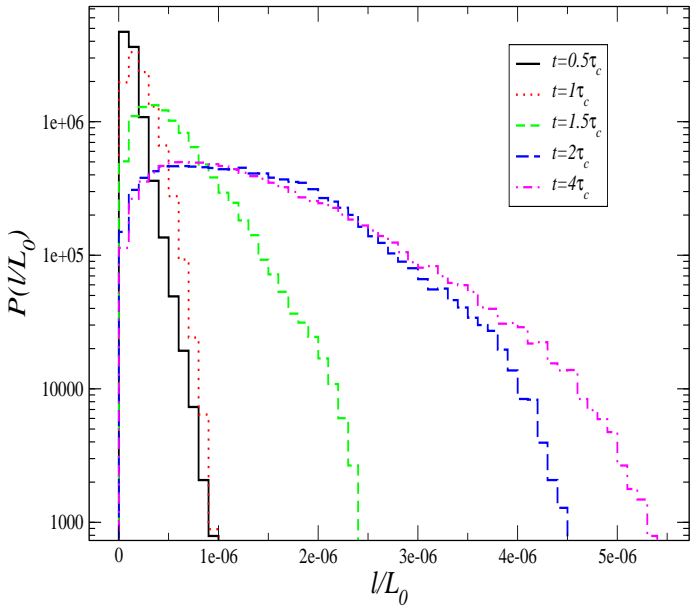}}
\par\centering
}
\caption{Distribution of the absolute values of the particle angular
momentum $\ell$ (in units of $L_0$) with respect to the center of
mass of bound particles $P(|\ell|)$, for $\alpha=1$ and $N=10^5$ at
different times before and after the collapse.}
\label{Fig_P_ell1} 
\end{figure}

In Fig.\ref{Fig_P_ell1} is shown the distribution $P(|\ell|)$ of the absolute
values of the particle angular momentum $\ell$ with respect to the
center of mass of bound particles, at a few different indicated 
times (before and after  the collapse). We observe, in line
with the results on the transverse velocities, that significant
angular momentum is generated already before the completion
of the collapse phase. Further  for $t>\tau_c$ $P(|\ell|)$ rapidly 
approaches the distribution characterizing the final virialized state.

\begin{figure}
\vspace{1cm}
{
\par\centering \resizebox*{9cm}{8cm}{\includegraphics*{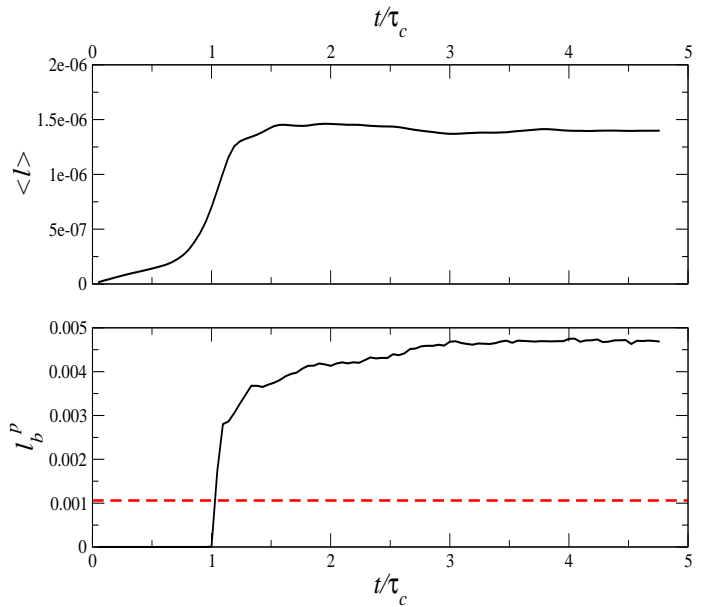}}
\par\centering
}
\caption{For the case $\alpha=1$, and $N=10^5$, the upper
panel shows the evolution of the average of the modulus
of the angular momenta of bound particles (with respect to 
their center of mass) as a function of time. The lower
panel shows the quantity in Eq.\ref{lbpn}, and the dashed
line the estimated value expected if the ejection 
were a random sampling of the total mass.}
\label{lambda2} 
\end{figure}

Shown in the upper panel of Figure \ref{lambda2} is the temporal
evolution of the average value of the modulus of the particle angular
momentum, for the bound particles (with respect to their center of
mass, in units of $L_0$). In line with what would be expected from the
plots of the transverse velocities, we observe a constant growth
during the collapse phase, followed by a very significant boost around
the time of maximal contraction, when the energy ejection occurs.
Thus we see the angular momentum generated by the collapse receives a
very significant boost in the final phase of the collapse. This is
further quantified by the lower panel of Figure \ref{lambda2} which
shows the evolution, as a function of time, of 
\be
\label{lbpn} 
l_b^p = \frac{L_b^p}{\sum_i |\ell|_i}
\ee
i.e. the modulus of the angular momentum of bound particles,
normalized to the sum of the moduli of the angular momenta of the same
particles. The dashed horizontal line corresponds to the value
$\sqrt{N_+}/N_- \approx f_+/\sqrt{N}$, which is the amplitude of the
bound angular momentum which would be expected if the ejection
operated as a simple random removal of particles without any
modification of their angular momentum.  We observe that the final
value is in fact at least several times larger. Thus, during the final
phase of collapse in which the large energy changes giving rise to
ejection occur, the ejected and bound particles exert a torque on one
another which very significantly amplifies their final (equal and
opposite) angular momenta.

\subsection{Results for different initial conditions (fixed $N$)} 

In the above we have focused, for simplicity, on the single case of
spherical initial conditions with $\alpha=1$, and we have shown
results for $N=10^5$. The qualitative behaviors we have discussed in
this case, for what concerns the generation of angular momentum, are
shared by all our other initial conditions.  Quantitatively there is a
variation of the final angular momentum, which depends notably on the
details of the mass ejection and the degree of symmetry breaking
during the collapse.  The precision of the simulations also varies
from case to case --- simulations for the spherical initial conditions
with $\alpha=0$ are particularly delicate as the collapse is the most
extreme in this case, with a singular behavior in the limit $N
\rightarrow \infty$.  Correspondingly in this case the measured
angular momentum is only a few times larger that the numerical error
in the conservation of the total angular momentum, while it is, as
shown above (see Fig. \ref{Fig1_alpha1_N1e5}) almost three orders of
magnitude for the case $\alpha=1$.

\begin{figure}
\vspace{1cm}
{
\par\centering \resizebox*{9cm}{8cm}{\includegraphics*{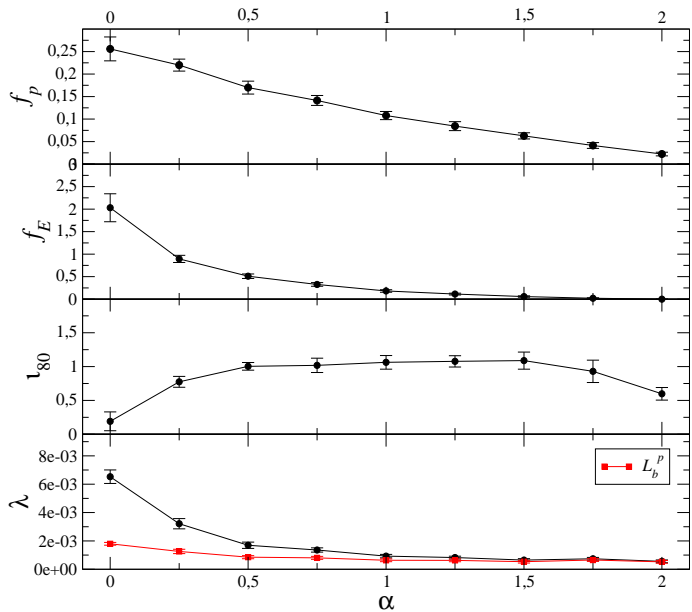}}
\par\centering
}
\caption{Final values of $f_+$, $\iota_{80}$ and $\lambda$ for the
  spherical initial conditions as a function of $\alpha$, for
  $N=10^4$.  The points are the average values for {$N_s=20$}
  realizations for each $\alpha \neq 0$, and for 47 realizations in
  the case $\alpha=0$; the indicated error bar is the corresponding
  standard deviation {$\sigma$} (for $\lambda$ we show the error
    on the mean {$\sigma/\sqrt{N_s}$}).  The lower panel
  also shows (red points, dashed) the angular momentum of the bound
  mass normalized to $L_0$. For the smallest values of $\alpha$ the
  angular momentum generated actually decreases, but, because of the
  increasing mass (and energy) ejection the characteristic angular
  momentum of the final bound mass decreases even faster.}
\label{Fig_alpha} 
\end{figure}

\begin{figure}
\vspace{1cm}
{
\par\centering \resizebox*{9cm}{8cm}{\includegraphics*{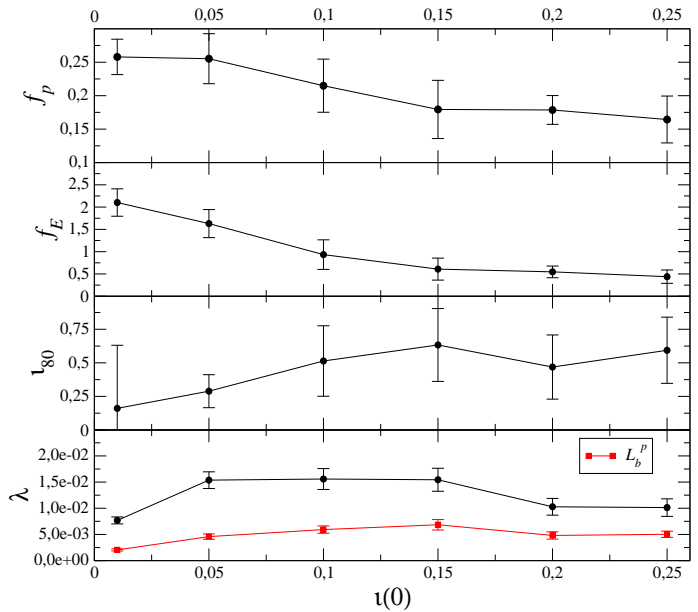}}
\par\centering
}
\caption{Final values of $f_+$, $\iota_{80}$ and $\lambda$ for
  ellipsoidal initial conditions, as a function of $\iota(0)$. The
  points are averages for 20 realizations, and the error bar the
  corresponding standard deviation {(for $\lambda$ we show the
    error on the mean)}. As in the previous figure the lower panel
  displays also (red points, dashed line) the final angular momentum
  normalized to $L_0$).}
\label{Fig_iota0} 
\end{figure}

Fig.\ref{Fig_alpha} shows the ``final" values of $f_+$ (top panel),
$\iota_{80}$ (third panel) and  $\lambda$ (bottom panel), i.e. at $t= 5 \tau_c$, 
for the spherical initial conditions  as a function of $\alpha$.
In addition, in the second panel, is shown also $f_E$,
the ``fraction of ejected energy'', defined by  
$f_E=(E_b-E_0)/E_0=-E_p/E_0$ where $E_p$
is (by energy conservation) the total ejected energy
in the frame in which the particles are initially at rest, 
i.e. the sum of the kinetic energy of the center
of mass of the bound mass and the kinetic energy of the 
ejected particles.
Fig.\ref{Fig_iota0} shows the same four quantities for 
the ellipsoidal initial conditions, as a function of $\iota(0)$. 

Comparing these two figures, the most striking result is that there is
an order of magnitude difference in the final angular momentum (as
characterized by the spin parameter $\lambda$) for most of the
spherical initial conditions with $\alpha \neq 0$, and the ellipsoidal
initial conditions with $\iota(0) \neq 0$, with a sharp interpolation
between the two classes around the case $\alpha=0$ (which indeed is
also the limit $\iota(0) \rightarrow 0$ of the ellipsoidal initial
conditions). In both figures we show also in the lower panel the final
angular momentum normalized to $L_0$. Comparing with $\lambda$, we see
that around $\alpha=0$, where the relaxation is most violent, there is
a significant difference.  This arises simply from the fact that there
is a considerable ejected mass, and energy, in this case: indeed the
relative amplitude of the two quantities can be expressed as
\begin{equation}
\frac{\lambda}{L_b^p}=\left(\frac{M}{M_b}\right)^{5/2}
\left(\frac{|E_b|}{|E_0|}\right)^{1/2}=
\frac{(1+f_E)^{1/2}}{(1-f_+)^{5/2}}\,.
\end{equation}
Thus the final spin includes an amplification which comes from the
change in the characteristic scale of angular momentum due to the mass
and energy ejection (which leads to a more bound, but less massive,
structure than the initial condition).

It is clear, however, from the two plots that this amplification from
the normalization is not the main factor explaining the much larger
final angular momenta from the ellipsoidal initial conditions. Rather
the main effect amplifying the spin in the ellipsoidal initial
conditions compared to the spherical ones comes clearly from the
essential difference in the initial conditions: the stronger initial
breaking of the spherical symmetry. This initial asymmetry both leads
to a more anisotropic collapse {compared to that in the spherical
case)} and, at the same time, the ejection of a comparable 
amount of mass {to that in the small $\alpha$ spherical
models}. As described in detail in
\cite{Benhaiem+SylosLabini_2015}, the particles on the stretched 
axis of the initial condition fall slightly later, and receive a large 
energy boost.
Note that it is only for $\iota_{80}(0)< 0.15$ that the final $\iota(t)$ 
 is linearly proportional to the initial one. Instead, for $\iota_{80}(0)
  > 0.15$ , substructures form during the collapse, leading to
  a substantial difference in the shape of the virialized
  structure \cite{Benhaiem+SylosLabini_2015}. Similarly the fraction
  of ejected particles decreases linearly with $\iota_{80}(0)$ only
  for $\iota_{80}(0)< 0.15$.

\begin{figure}
\vspace{1cm} { \par\centering
  \resizebox*{9cm}{8cm}{\includegraphics*{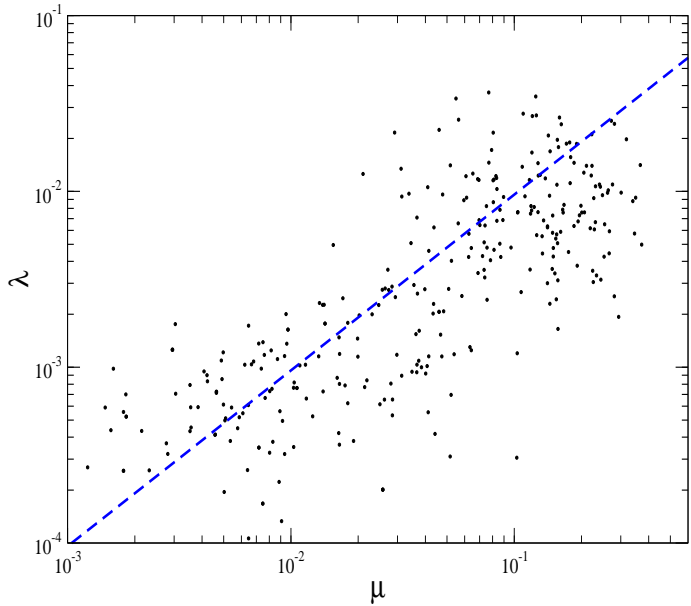}}
  \par\centering }
\caption{Final spin parameter $\lambda$, and the modulus of final total linear 
momentum of the bound particles $\mu$ , in all our (322) realizations of
initial conditions (both spherical and ellipsoidal) with $N=10^4$.
$\mu$ is normalized in units in which $GM^3/R_0=1$.   
The dashed line is $ \lambda \propto \mu$. }
\label{L-Pcorrelation} 
\end{figure}

Besides the angular momentum itself, the generation of a net relative
motion of the center of masses of the final bound and ejected mass is
also a direct measure of symmetry breaking. We would thus expect that
we might find a similar trend in the final linear momentum of the
ejected (or bound) particles. Shown in Fig.  \ref{L-Pcorrelation} is a
plot, for all our simulations with $N=10^4$ particles, of both
$\lambda$ and $\mu$, the modulus of the ``final" linear momentum of
the ejected particles (equal and opposite to that of the bound mass up
to a numerical precision which is smaller by several orders of
magnitude in all cases.) We observe that there is indeed a clear
correlation between the two quantities. Further it appears highly
consistent with a roughly linear relation, as one might expect, given
that they are both indirect measures of an initially small spherical
symmetry breaking.

\begin{figure}
\vspace{1cm} { \par\centering
  \resizebox*{9cm}{8cm}{\includegraphics*{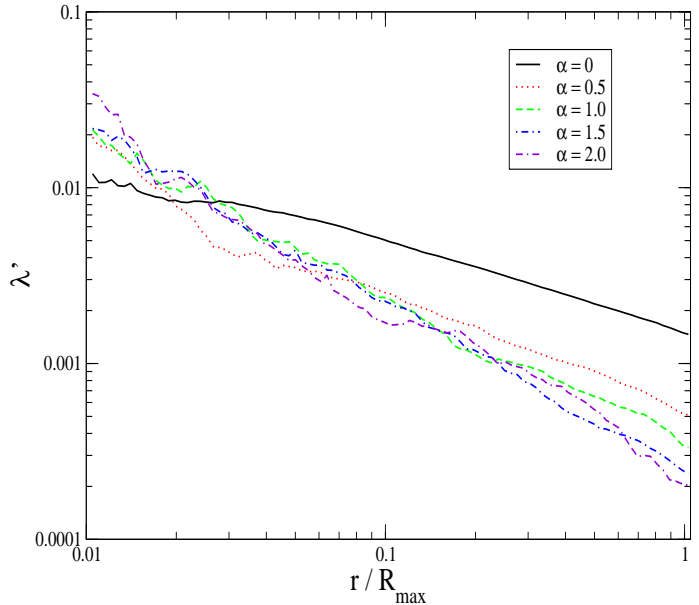}}
  \par\centering }
\caption{Spin parameter $\lambda'(r)$ (see Eq.\ref{lambdap}),
    averaged over 20 realizations, for $\alpha=0,0.5,1,1.5,2$ and
    $N=10^5$. The distance is normalized to the structure size
    $R_{max}$. }
\label{Ang_mom_profile} 
\end{figure}

We have studied finally also the scale dependence of the normalized 
spin parameter defined as \citep{Bullock_2001}
\be
\label{lambdap} 
\lambda'(r) = \frac{L_b^p(r)}{\sqrt{2 G M(r)^3 r}}
\ee
where $L_b^p(r)$ is the angular momentum of bound particles with
distance $<r$ from their center of mass and $M(r)$ is the mass
enclosed in a sphere of radius $r$. Results for five different sets of
simulations with $\alpha=0,0.5,1,1.5,2$ and $N=10^5$ are shown in
Fig.\ref{Ang_mom_profile}.  The profiles are similar in all cases, but
their amplitude depends on $\alpha$ as highlighted in
Fig.\ref{Fig_alpha}. The fact that $\lambda'(r)$ is larger for
$\alpha=0$ than for $\alpha>0$ is related to the fact orbits are
slightly more radial in the former case.

\subsection{Correlation of angular momentum with final spatial distribution} 

As analyzed in detail in \cite{Benhaiem+SylosLabini_2015,SylosLabini+Benhaiem+Joyce_2015}
there is, for both classes of initial conditions, a correlation between the direction
in which mass is ejected and the longest axis of both the initial
condition and the final bound mass. These latter two are strongly
correlated because it is the growth of the initial fluctuations breaking
spherical symmetry, amplified by the mass ejection, which leads
to the final spatial asymmetry. The fact that the ejection of mass
is amplified along these same directions is because they 
are the particles which fall latest and pick up as a result a 
large energy kick. Given that we have seen that most of the
final angular momentum is generated at the time of ejection, 
we would expect that it would preferably be aligned orthogonal
to the preferential axis for ejection. 

\begin{figure}
\vspace{1cm} { \par\centering
  \resizebox*{9cm}{8cm}{\includegraphics*{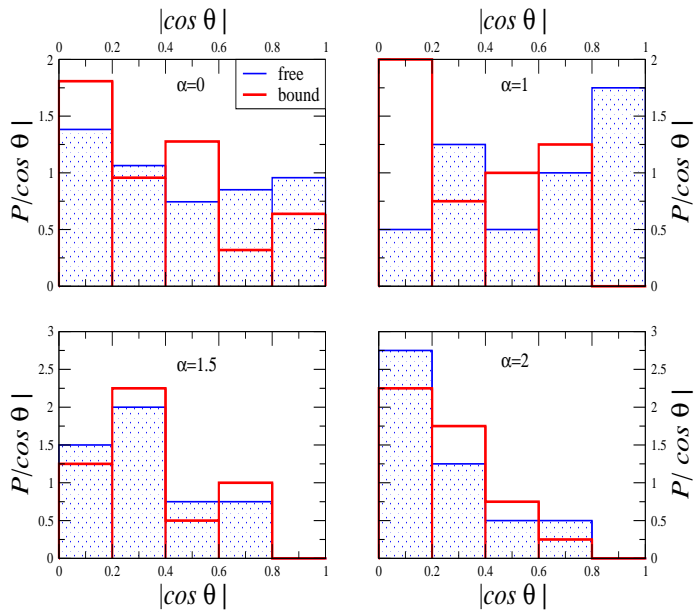}}
  \par\centering }
\caption{Histogram of the cosine of the angle, well after the
  collapse, between $\vec{L}_b^p$ and the eigenvector corresponding to
  the longest principal axis of the final bound mass (blank
    histograms), and of the ejected mass (filled histograms),
  for 20 realizations of spherical initial conditions for different
  $\alpha$ indicated in the caption.}
\label{Fig_alpha_costheta} 
\end{figure}

\begin{figure}
\vspace{1cm} { \par\centering
  \resizebox*{9cm}{8cm}{\includegraphics*{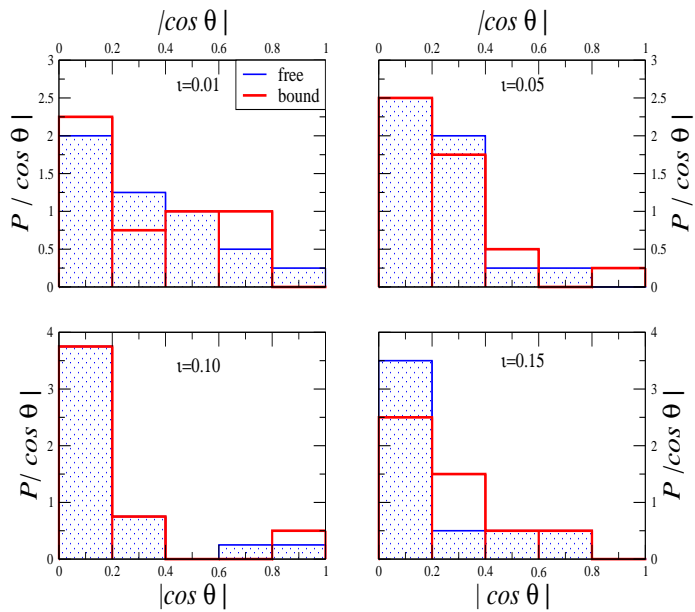}}
  \par\centering }
\caption{Pdf of the cosine of the angle, well after the collapse,
  between $\vec{L}_b^p$ and the eigenvector corresponding to the
  longest principal axis of the final bound (blank histograms),
  and of the ejected mass (filled histograms), for 20
  realizations of ellipsoidal initial conditions for different
  $\iota(0)$ indicated in the caption.}
\label{Fig_iota0_costheta} 
\end{figure}

Shown in Figs. \ref{Fig_alpha_costheta} and \ref{Fig_iota0_costheta}
are histograms, for 20 realizations of each indicated initial
condition, of the modulus of the cosine of the angle between the final
angular momentum of the bound mass, $\vec{L}_b^p$, and the eigenvector
corresponding to the longest axis of the final bound mass (blank
  histograms), and the final ejected mass (filled histograms)
For the ellipsoidal initial conditions - which very strongly single
out an initial preferred axis which remains strongly correlated with
the final longest axis - the anticipated correlation is very clearly
present. In the spherical initial conditions, the correlation is
weaker but nevertheless visible, except perhaps in the case
$\alpha=0$. Indeed in this latter case, as discussed in
\cite{SylosLabini+Benhaiem+Joyce_2015}, the symmetry breaking
in the final state is in fact much weaker, and the correlation between
the final and initial asymmetry likewise.  This is a result of the
fact that in this case, in which the fall time of all particles is the
same in the limit $N \rightarrow \infty$, there is a much weaker
correlation between the initial radial position of a particle and the
energy change it undergoes in the violent collapse.

\subsection{Dependence on $N$} 

The $N$ dependence of the angular momentum generated by mass ejection
has been explored in \cite{worrakitpoonpon_2014} for the case
$\alpha=0$, and {evidence for a monotonic decrease 
$\sim N^{-\beta}$, with a best fit around $\beta=1/3$, was found}.
Inspection of the
results we have reported above, for simulations with $N=10^4$ and
$N=10^5$, are consistent with this finding for this case, while for
initial conditions with $\alpha >0$, {an apparent decrease of 
the average $L_p^b$ as $N$ grows is also observed}. 
To explore this further, we have performed a more detailed study of the 
case $\alpha=1$, performing a larger number of simulations for a greater 
number of values of $N$ in the range $10^3$ to $10^5$. We have in 
this case also evolved each simulation for a longer time to study 
the stability of our ``final'' measures of the different quantities, and 
to check for the possible role of finite $N$ effects.

\begin{figure}
\vspace{1cm}
{
\par\centering \resizebox*{9cm}{8cm}{\includegraphics*{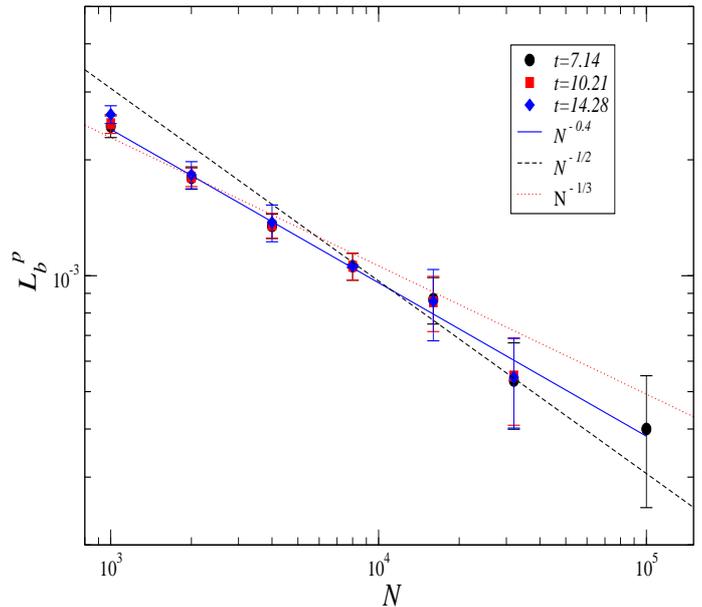}}
\par\centering
}
\caption{Logarithm of the final angular momentum of bound virialized
  mass $L_b^p$ (in units of $L_0$) as a function of particle number
  $N$, for spherical initial conditions with $\alpha=1$. The point
  for the cases $N=10^5$ correspond to the results
  reported above, at $t=5$, while for the other $N$ there is data at
 three times, extending to $t \approx 14$.  Each point is an
 average over $M$ realizations, with $M=90$ for $N=1000$, $M=50$ for
  $N=2000$, $M=40$ for $N=4000$, $M=30$ for $N=8000$, $M=6$ for
  $N=16000$, $M=4$ for $N=32000$, and $M=4$ for $N=10^5$. The error
 bars indicate the {corresponding estimated error on the mean}.  
 The continuous line  {($\propto N^{-0.4}$) is a best-fit to the data, 
 while the dashed lines corresponds to $N^{-1/2}$ and
 $N^{-1/3}$.}
 }
\label{Ndependence-alpha1-cold} 
\end{figure}

Shown in Fig. \ref{Ndependence-alpha1-cold} is the result for the
measured angular momentum $L_b^p$, at the indicated times, in a series
of simulations for the different indicated values of $N$.  As
anticipated, in the cases where different (and longer) times have been
analyzed the measured angular momentum shows no significant dependence
on the time: after the few dynamical time there is negligible
interaction between the ejected and bound mass and the angular
momentum of each are constant.  As a function of $N$, on the other
hand, there is a clear monotonic decrease, well fitted by a power law
dependence, $L_b^p \propto N^{-\gamma}$, with 
$\gamma \approx 0.4$ 
{The dashed lines indicate, for comparison, the behaviors 
$N^{-1/2}$ and $N^{-1/3}$.}

\begin{figure}
\vspace{1cm}
{
\par\centering \resizebox*{9cm}{8cm}{\includegraphics*{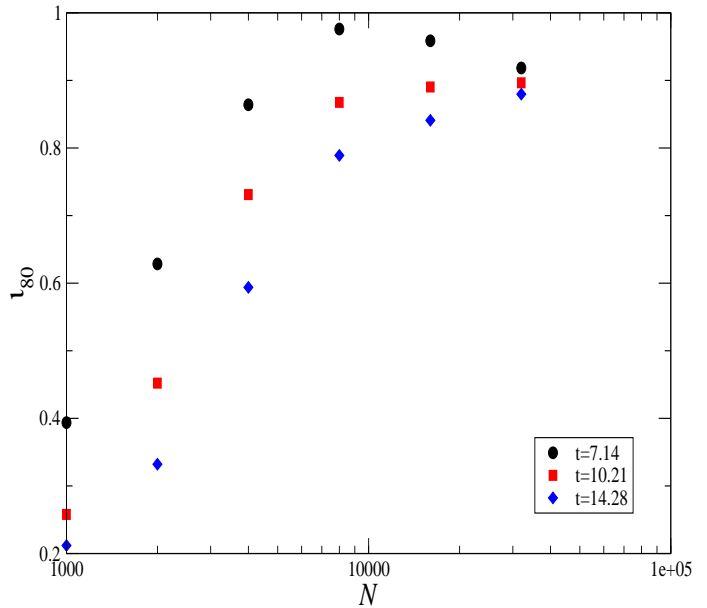}}
\par\centering
}
\caption{Flatness ratio $\iota_{80}$ as a function of particle number $N$, 
for the same simulations in the previous figure ($\alpha=1$) for the 
indicated values of $N$. For this quantity the observed $N$ dependence 
is due to two body relaxation which make the system evolve in time towards 
a more spherical quasi-equilibrium.}
\label{Ndependence-iota80-alpha=1} 
\end{figure}

Shown in Fig. \ref{Ndependence-iota80-alpha=1} is the result for the
same data as in the previous figure, and for the different $N$ shown,
for the flatness ratio $\iota_{80}$.  In contrast we see in this case
that the value measured, for all but the largest value of $N$, shows
visible dependence on time. More specifically the bound mass evolves
towards a more spherical distribution in time, at a rate which is
clearly faster for the smaller $N$. Such an $N$ dependence of the time
scale of the evolution of the virialized structure is clearly
qualitatively coherent with collisional relaxation, and indeed a
quantitative analysis of such behavior from similar initial
conditions in \cite{theis+spurzem_1999} shows that the
timescale of this evolution is indeed in agreement with that predicted
by two-body collisionality.  At the very largest $N$, on the other
hand, the result appears to converge to a time independent value.
{In the small range of (larger) $N$ in which this time-dependence 
due to two body collisionality is small the final value appears to be 
consistent with an $N$-independent behavior, but clearly
extrapolation to larger $N$ would be needed to reach a firmer
conclusion. We underline that, in contrast, we do not have 
this problem for the determination above of the scaling
with $N$ of $L_p^b$ as this quantity is essentially
frozen, as we have seen, at about $t \approx 2$,
and thus not modified by the two body collisionality
at longer times.}
We note that  \cite{joyce_etal_2009} report tests for 
collisionality during the collapse phase, for the very
extreme case of a flat ($\alpha=0$) profile, and conclude, in line with 
the study of \cite{Aarseth_etal_1988}, that for  $N>10^3$ collisional 
effects during the collapse phase are indeed negligible.
This result, consistent with theoretical prediction for the collisional 
time scale,  is borne out by the observation that the
evolution of the macroscopic features of the collapse 
--- potential energy, viral ratio, size of structure, and in
particular the fraction of the mass ejected --- are 
stable when $N$ increases, and likewise when 
the gravitational force softening is varied, provided it
is sufficiently small to resolve the mean gravitational field.

Given that, as we have emphasized, the breaking of spherical symmetry
  in the initial conditions is due to finite $N$ fluctuations, the
  fact we find that the final angular momentum $L_b^p$ decreases as
  $N$ increases, and in principle goes to zero as $N \rightarrow
  \infty$, is what one would naively expect. A similar decrease
  in amplitude, {with an exponent of about the same value}, 
  has been observed for the case $\alpha=0$ in \cite{worrakitpoonpon_2014}. 
  In our simulations of
  initial conditions from elliptical conditions we have not been able
  to detect, within the significant scatter of different realizations
  at given $N$, a clear trend for $L_b^p$ to decrease as $N$
  increases. This is probably because of a weaker $N$ dependence in
  this case, as the spherical symmetry is ``mostly" broken by
  $\iota(0) \neq 0$ which does not depend on $N$.  In contrast the
  lack of evidence for any $N$ dependence in the final $\iota$ ---
  once $N$ is sufficiently large so that collisional relaxation plays
  no role on the time scales probed --- may be indicative of the true
  physical behavior: in the limit of exact spherical symmetry of an
  exactly cold system, it is expected that symmetry be broken due to
  so-called radial orbit instability.  In future work we will study
 more fully the $N$ dependence of both the breaking of symmetry, 
 and that of the angular momentum of the ejected particles, and 
their connection to one another.

\section{Discussion and conclusions} 
\label{diss_concl}

The origin of the angular momentum in astrophysical structures ---
notably that measured observationally in galaxies ---- is a
fascinating open problem in astrophysics and cosmology.  In this
article we have discussed and studied a way, different to commonly
considered ones such as tidal torques, in which angular momentum can
be generated by self-gravity only.  It can occur in principle for
initial conditions which evolve sufficiently violently --- with large
variations of the gravitational mean field --- so that mass is ejected
during relaxation towards a virialized state. We have shown using
numerical simulations for two simple broad classes of initial
conditions of this kind that significant angular momentum can indeed
be generated in this way, orders of magnitude larger in most cases
that the angular momentum error due to numerics.

We have, for illustrative purposes, and for simplicity, considered
initial conditions very close to spherical symmetry. Indeed in all our
initial conditions rotational symmetry is only broken fully by the
Poissonian density fluctuations associated with the finite number of
particles. Interestingly we find that, despite the ``artificiality" of
our initial conditions, which are not motivated by any detailed
specific physical model, we obtain angular momenta which are of
comparable size --- only smaller by a factor of two in some cases ---
{ than those typically estimated for galaxies}.  Indeed we have
obtained values of the normalized ``spin parameter" up to $\lambda
\sim 0.02$, while, for example, \cite{hernandez2007empirical} have
estimated its average value in the $11597$ spirals and elliptical
galaxies observed by the Sloan Digital Sky Survey to be
$\lambda_0=0.04 \pm 0.005$, (and standard deviation
$\sigma_{\lambda}=0.51 \pm 0.05$).  While obviously galaxy formation
in particular involves much more complicated {non-gravitational
  processes than described by the simple models we have considered
  here, and further our specific initial conditions are quite ad hoc
  (and not directly motivated by a cosmological model) it is
  intriguing that the order of magnitude of the values we obtain are
  in line with those observed.}

{We note that studies of dark matter halos in cosmological N-body
  simulations (see e.g.
  \cite{vitvitska2002origin,Bullock_2001,bett2007spin}) give
  rise to similar values of spin parameter ($\lambda \approx 0.04$),
  i.e. of the same order as those observed in elliptical galaxies, and
  as those we find in our simulations in certain cases.  As dark
  matter halos of standard cold dark matter cosmologies form through a
  hierarchical process rather than in the kind of monolithic collapse
  we have studied, the mechanism we have studied has likely no
  relevance in this particular context. Indeed it is the strong
  violence of the collapse leading to mass ejection which is crucial,
  and the formation of halos in a cold dark matter cosmology setting
  is not of this kind. Nevertheless, as discussed in
  \cite{carucci_etal_2014, samsing_2015}, significant mass ejection
  can occur in a cosmological setting when halos merge and it would be
  interesting to investigate whether this could lead also to
  generation of angular momentum at a significant level, compared to
  the processes of mass accretion, for example, which has been argued
  in \cite{vitvitska2002origin} to account for the spin of dark matter
  halos.}

Just as in cosmological simulations we underline that
we are simulating essentially the collisionless regime of the gravitational 
dynamics (apart from the two body collisional effects which 
we observed to be present at longer times in smaller $N$
simulations). In so far as this
is true, the particle number $N$ is relevant, in principle, in the properties 
of the final state just because it fixes the amplitude of the initial
fluctuations. Indeed these fluctuations are, for our initial conditions
in this paper, simply Poissonian, and thus their amplitude at any 
scale (and all their statistical properties) are regulated by this 
single parameter. Alternatively one could consider setting up an initial 
particle with the same average density profile but with statistical 
fluctuations with non-trivial correlation, described for example by 
non-trivial two-point correlation properties. In this case one could then
in principle vary $N$ while keeping the relevant 
fluctuations fixed and obtain results independent of $N$.
As the fluctuations play an essential role in the symmetry breaking, 
we expect that the detailed nature of such initial fluctuations 
may have an  impact on the properties of the final structure.
We postpone a detailed study of this interesting issue to future work.
We will also explore whether cold but more irregular/clumpy
initial conditions can produce values of the spin of the order
of magnitude to those observed for galaxies. Further we will
explore whether the kind of correlation we have discussed briefly 
in the last part of the paper --- between the
shape of the asymmetric virialized structure and its angular momentum
--- could be used to find observational evidence for or against the
role of very violent relaxation in generating the structure of
galaxies.

\bigskip 

Numerical simulations have been run on the Cineca PLX cluster (project
ISCRA BSS-GC), and on the HPC
resources of The Institute for Scientific Computing and Simulation
financed by Region Ile de France and the project Equip@Meso (reference
ANR-10-EQPX- 29-01) overseen by the French National Research Agency
(ANR) as part of the Investissements d'Avenir program.

\bibliographystyle{mn2e}

\begin{thebibliography}{32}
\expandafter\ifx\csname natexlab\endcsname\relax\def\natexlab#1{#1}\fi

\bibitem[{Aarseth, Lin \& Papaloizou(1988)Aarseth, Lin, \&
  Papaloizou}]{Aarseth_etal_1988}
Aarseth S., Lin D., Papaloizou J., 1988, Astrophys. J., 324, 288

\bibitem[{Aguilar \& Merritt(1990)}]{aguilar+merritt_1990}
Aguilar L., Merritt D., 1990, Astrophys. J., 354, 73

\bibitem[{{Antonov}(1961)}]{antonov_1961}
{Antonov} V.~A., 1961, Soviet Ast., 4, 859

\bibitem[{{Barnes}, {Lanzel} \& {Williams}(2009){Barnes}, {Lanzel}, \&
  {Williams}}]{barnes_etal_2009}
{Barnes} E.~I., {Lanzel} P.~A., {Williams} L.~L.~R., 2009, Astrophys. J., 704,
  372

\bibitem[{{Bellazzini} {et~al}\mbox{.}(2012){Bellazzini}, {Bragaglia},
  {Carretta}, {Gratton}, {Lucatello}, {Catanzaro}, \& {Leone}}]{globclu3}
{Bellazzini} M., {Bragaglia} A., {Carretta} E., {Gratton} R.~G., {Lucatello}
  S., {Catanzaro} G., {Leone} F., 2012, Astron.Astrophys., 538, A18

\bibitem[{{Benhaiem} \& {Sylos Labini}(2015)}]{Benhaiem+SylosLabini_2015}
{Benhaiem} D., {Sylos Labini} F., 2015, Mon.Not.R.Astron.Soc., 448, 2634

\bibitem[{Bett {et~al}\mbox{.}(2007)Bett, Eke, Frenk, Jenkins, Helly, \&
  Navarro}]{bett2007spin}
Bett P., Eke V., Frenk C.~S., Jenkins A., Helly J., Navarro J., 2007, Monthly
  Notices of the Royal Astronomical Society, 376, 215

\bibitem[{Boily, Athanassoula \& Kroupa(2002)Boily, Athanassoula, \&
  Kroupa}]{Boily_etal_2002}
Boily C., Athanassoula E., Kroupa P., 2002, Mon. Not. R. Astr. Soc., 332, 971

\bibitem[{{Boily} \& {Athanassoula}(2006)}]{boily+athanassoula_2006}
{Boily} C.~M., {Athanassoula} E., 2006, Mon. Not. R. Astr. Soc., 369, 608

\bibitem[{{Bullock} {et~al}\mbox{.}(2001){Bullock}, {Dekel}, {Kolatt},
  {Kravtsov}, {Klypin}, {Porciani}, \& {Primack}}]{Bullock_2001}
{Bullock} J.~S., {Dekel} A., {Kolatt} T.~S., {Kravtsov} A.~V., {Klypin} A.~A.,
  {Porciani} C., {Primack} J.~R., 2001, Astrophys.J, 555, 240


\bibitem[{{Carucci} {et~al}\mbox{.}(2014){Carucci}, {Sparre}, {Hansen}, \&
  {Joyce}}]{carucci_etal_2014}
{Carucci} I.~P., {Sparre} M., {Hansen} S.~H., {Joyce} M., 2014, JCAP, 6, 57

\bibitem[{{Fridman} {et~al}\mbox{.}(1984){Fridman}, {Polyachenko}, {Aries}, \&
  {Poliakoff}}]{Fridman+Polyachenko_etal_1984}
{Fridman} A.~M., {Polyachenko} V.~L.
1984, {Physics of gravitating systems. I. Equilibrium and
    stability.} Springer Verlag

\bibitem[{{H{\'e}nault-Brunet} {et~al}\mbox{.}(2012){H{\'e}nault-Brunet},
  {Gieles}, {Evans}, {Sana}, {Bastian}, {Ma{\'{\i}}z Apell{\'a}niz}, {Taylor},
  {Markova}, {Bressert}, {de Koter}, \& {van Loon}}]{globclu1}
{H{\'e}nault-Brunet} V. {et~al.}, 2012, Astron.Astrophys., 545, L1

\bibitem[{{Henon}(1973)}]{Henon_1973}
{Henon} M., 1973, Ann. Astrophys., 24, 229

\bibitem[{Hernandez {et~al}\mbox{.}(2007)Hernandez, Park, Cervantes-Sodi, \&
  Choi}]{hernandez2007empirical}
Hernandez X., Park C., Cervantes-Sodi B., Choi Y.-Y., 2007, Monthly Notices of
  the Royal Astronomical Society, 375, 163

\bibitem[{{Joyce}, {Marcos} \& {Sylos Labini}(2009){Joyce}, {Marcos},
    \& {Sylos Labini}}]{joyce_etal_2009} {Joyce} M., {Marcos} B.,
  {Sylos Labini} F., 2009, Mon. Not. R. Astron. Soc., 397, 775


\bibitem[\protect\citeauthoryear{Joyce \& Sylos Labini}{Joyce \& Sylos
    Labini}{2012}]{jsl13} Joyce M., Sylos Labini F., 2013,
  Mon.Not.R.Astr.Soc.,  429, 1088


\bibitem[{{Kacharov} {et~al}\mbox{.}(2014){Kacharov}, {Bianchini}, {Koch},
  {Frank}, {Martin}, {van de Ven}, {Puzia}, {McDonald}, {Johnson}, \&
  {Zijlstra}}]{globclu2}
{Kacharov} N. {et~al.}, 2014, Astron.Astrophys., 567, A69

\bibitem[{{Knebe} \& {Power}(2008)}]{knebe2008}
{Knebe} A., {Power} C., 2008, Astrophys.J., 678, 621

\bibitem[{{Lin}, {Mestel} \& {Shu}(1965){Lin}, {Mestel}, \&
  {Shu}}]{Lin_Mestel_Shu_1965}
{Lin} C.~C., {Mestel} L., {Shu} F.~H., 1965, Astrophys. J., 142, 1431

\bibitem[{{Merritt} \& {Aguilar}(1985)}]{merritt+aguilar_1985}
{Merritt} D., {Aguilar} L.~A., 1985, Mon. Not. R. Ast. Soc, 217, 787

\bibitem[{{Peebles}(1969)}]{Peebles_1969}
{Peebles} P.~J.~E., 1969, Astrophys.J., 155, 393

\bibitem[{{Romanowsky} \& {Fall}(2012)}]{romanowsky+fall_2012}
{Romanowsky} A.~J., {Fall} S.~M., 2012, Astrophys. J. Supp, 203, 17

\bibitem[{{Samsing}(2015)}]{samsing_2015}
{Samsing} J., 2015, Astrophys.J., 799, 145

\bibitem[{Springel(2005)}]{springel_2005}
Springel V., 2005, Mon. Not. R. Ast. Soc., 364, 1105

\bibitem[{{Sylos Labini}(2012)}]{syloslabini_2012}
{Sylos Labini} F., 2012, Mon. Not. R. Astron. Soc., 423, 1610

\bibitem[{{Sylos Labini}(2013)}]{syloslabini_2013}
{Sylos Labini} F., 2013, Mon. Not. R. Astron. Soc., 429, 679

\bibitem[\protect\citeauthoryear{Sylos Labini}{Sylos
    Labini}{2012}]{sl13} Sylos Labini F., 2013, Astronomy \&
  Astrophysics, 552, A36

\bibitem[{{Sylos Labini}, {Benhaiem} \& {Joyce}(2015){Sylos Labini},
  {Benhaiem}, \& {Joyce}}]{SylosLabini+Benhaiem+Joyce_2015}
{Sylos Labini} F., {Benhaiem} D., {Joyce} M., 2015, Mon.Not.R.Astron.Soc., 449,
  4458

\bibitem[{Theis \& Spurzem(1999)}]{theis+spurzem_1999}
Theis C., Spurzem R., 1999, Astron. Astrophys., 341, 361

\bibitem[{van Albada(1982)}]{vanalbada_1982}
van Albada T., 1982, Mon. Not. R. Astr. Soc., 201, 939

\bibitem[{Vitvitska {et~al}\mbox{.}(2002)Vitvitska, Klypin, Kravtsov, Wechsler,
  Primack, \& Bullock}]{vitvitska2002origin}
Vitvitska M., Klypin A.~A., Kravtsov A.~V., Wechsler R.~H., Primack J.~R.,
  Bullock J.~S., 2002, The Astrophysical Journal, 581, 799

\bibitem[{{Worrakitpoonpon}(2015)}]{worrakitpoonpon_2014}
{Worrakitpoonpon} T., 2015, Mon. Not. R. Astr. Soc., 466, 1335

\end{thebibliography}

\end{document}